\documentclass[sigconf,screen]{acmart}

\AtBeginDocument{%
  \providecommand\BibTeX{{%
    \normalfont B\kern-0.5em{\scshape i\kern-0.25em b}\kern-0.8em\TeX}}}
    
\usepackage{amsmath,amsfonts}
\usepackage{microtype}
\usepackage{bm}
\usepackage{graphicx}
\usepackage{textcomp}
\usepackage{color,colortbl}
\usepackage{multirow}
\usepackage{float}
\usepackage{tabularx}
\usepackage{booktabs}
\usepackage[OT1]{fontenc}
\usepackage{wrapfig}
\usepackage{subcaption}
\usepackage{enumitem}
\allowdisplaybreaks[1]
\usepackage{siunitx}

\usepackage{makecell}
\usepackage{adjustbox}
\usepackage[nameinlink]{cleveref}
\usepackage{tikz}
\usepackage{tcolorbox}

\usepackage{xltabular}

\newcommand{\circtext}[1]{{\textcircled{\small{#1}}}}

\usepackage{mathcommand}

\renewtextcommand\_{\textunderscore\allowbreak}

\newcommand{\ours}[0]{\texorpdfstring{\textit{DéjàVu}}{DéjàVu}}

\newcommand{\ccb}[0]{\textit{compA}}

\newcommand{\sys}[1]{\texorpdfstring{\textit{sys#1}}{system #1}}

\newcommand{\rnum}[1]{\texorpdfstring{\uppercase\expandafter{\romannumeral #1}}{#1}}
\newcommand{\ds}[1]{\texorpdfstring{$\mathcal{#1}$}{dataset #1}}

\newcommand{\thickhline}{
    \Xhline{2\arrayrulewidth}
}

\AtBeginDocument{%
\crefformat{equation}{(#2#1#3)}
\crefformat{section}{#2\S~#1#3}
\crefformat{subsection}{#2\S~#1#3}
\crefformat{subsubsection}{#2\S~#1#3}
\crefformat{algorithm}{#2Algorithm~#1#3}
\crefformat{figure}{#2Fig.~#1#3}
\crefformat{subfigure}{#2Fig.~#1#3}
\crefformat{table}{#2Table~#1#3}

\Crefformat{equation}{(#2#1#3)}
\Crefformat{section}{#2\S~#1#3}
\Crefformat{subsection}{#2\S~#1#3}
\Crefformat{subsubsection}{#2\S~#1#3}
\Crefformat{algorithm}{#2Algorithm~#1#3}
\Crefformat{figure}{#2Fig.~#1#3}
\Crefformat{subfigure}{#2Fig.~#1#3}
\Crefformat{table}{#2Table~#1#3}
}

\newtheoremstyle{compact}{2pt}{2pt}{}{}{\bfseries}{.}{.5em}{}
\theoremstyle{compact}
\newtheorem{definition}{Definition}

\setcopyright{rightsretained}
\acmPrice{}
\acmDOI{10.1145/3540250.3549092}
\acmYear{2022}
\copyrightyear{2022}
\acmSubmissionID{fse22main-p98-p}
\acmISBN{978-1-4503-9413-0/22/11}
\acmConference[ESEC/FSE '22]{Proceedings of the 30th ACM Joint European Software Engineering Conference and Symposium on the Foundations of Software Engineering}{November 14--18, 2022}{Singapore, Singapore}
\acmBooktitle{Proceedings of the 30th ACM Joint European Software Engineering Conference and Symposium on the Foundations of Software Engineering (ESEC/FSE '22), November 14--18, 2022, Singapore, Singapore}

\begin{document}

\begin{abstract}
Fault localization is challenging in an online service system due to its monitoring data's large volume and variety and complex dependencies across/within its components (e.g., services or databases). 
Furthermore, engineers require fault localization solutions to be actionable and interpretable, which existing research approaches cannot satisfy. 
Therefore, the common industry practice is that, for a specific online service system, its experienced engineers focus on localization for recurring failures based on the knowledge accumulated about the system and historical failures.
More specifically, 1) they can identify the underlying root causes and take mitigation actions when pinpointing a group of indicative metrics on the faulty component; 2) their diagnosis knowledge is roughly based on how one failure might affect the components in the whole system.

Although the above common practice is actionable and interpretable, it is largely manual, thus slow and sometimes inaccurate.
In this paper, we aim to automate this practice through machine learning.
That is, we propose an actionable and interpretable fault localization approach, \ours{}, for recurring failures in online service systems.
For a specific online service system, \ours{} takes historical failures and dependencies in the system as input and trains a localization model offline; 
for an incoming failure, the trained model online recommends where the failure occurs (i.e., the faulty components) and which kind of failure occurs (i.e., the indicative group of metrics) (thus actionable), which are further interpreted both globally and locally (thus interpretable).
Based on the evaluation on \num{601} failures from three production systems and one open-source benchmark, in less than one second, \ours{} can rank the ground truths at 1.66$\sim$5.03-\textit{th} among a long candidate list on average, outperforming baselines by 54.52\%.
\end{abstract}

\title{Actionable and Interpretable Fault Localization for~Recurring~Failures in Online Service Systems}

\author{Zeyan Li}
\author{Nengwen Zhao}
\author{Mingjie Li}
\author{Xianglin Lu}
\affiliation{
    \institution{Tsinghua University}
    \city{Beijing}
    \country{China}
}

\author{Lixin Wang}
\author{Dongdong Chang }
\affiliation{
    \institution{China Construction Bank}
    \city{Beijing}
    \country{China}
}

\author{Xiaohui Nie}
\author{Li Cao}
\author{Wenchi Zhang}
\author{Kaixin Sui}
\affiliation{
    \institution{BizSeer}
    \city{Beijing}
    \country{China}
}

\author{Yanhua Wang}
\author{Xu Du}
\author{Guoqiang Duan}
\affiliation{
    \institution{China Construction Bank}
    \city{Beijing}
    \country{China}
}

\author{Dan Pei}
\authornote{Dan Pei is the corresponding author.}
\affiliation{
    \institution{Tsinghua University}
    \city{Beijing}
    \country{China}
}

\renewcommand{\shortauthors}{Z. Li, N. Zhao, M. Li, X. Lu, L. Wang, D. Chang, X. Nie, L. Cao, W. Zhang, K. Sui, Y. Wang, X. Du, G. Duan, D. Pei}



\maketitle

\section{Introduction}
\label{sec:introduction}
Recently, online service systems (e.g., online shopping platforms or E-banks) have gradually replaced traditional software systems and play an indispensable part in our daily life ~\cite{chen2019continuous,chen2019empirical,zhang2019robust,chen2022adaptive}.
Though tremendous effort has been devoted to software service maintenance (e.g., various metrics, such as average response time or memory usage, are closely monitored on a 24$\times$7 basis~\cite{chen2022adaptive}), failures are inevitable due to the large scale and complexity, causing huge economic loss and user dissatisfaction \cite{lin2016log,pham2017failure,mariani2020predicting,brandon2020graphbased}.

To enable engineers to resolve failures efficiently, fault localization is at the core of software maintenance for online service systems~\cite{chen2020intelligent,dang2019aiops,lou2013software}.
However, existing approaches mainly focus on unactionable fault levels, e.g., individual metrics~\cite{liu2019fluxrank,wu2021identifying,kim2013root} or components~\cite{zhou2019latent,pham2017failure,gan2021sagea,liu2020microhecl}, which, respectively, can be too fine-grained (e.g., it is hard to tell what exactly happened if only \textit{memory usage} is localized) or coarse-grained (e.g., diagnosing faulty components is still challenging).
\textit{
To be actionable, we aim to inform engineers where the failure occurs (i.e., the faulty component) and which kind of failure occurs (e.g., memory leak).
}
For convenience, we name the combination of a specific kind of failure and a location as a \textit{failure unit}.
More specifically, monitoring metrics serve as the most direct signals to the underlying failures~\cite{chen2022adaptive}, and different kinds of failures (thus the corresponding mitigation actions) can be indicated by different groups of metrics on the faulty components~\cite{thalheim2017sieve,chen2022adaptive}.
For example, a combination of high memory usage and high \#requests indicates insufficient memory caused by bursting requests rather than a memory leak.
In summary, we aim to recommend the faulty components and the corresponding indicative metric groups to engineers, i.e., the faulty failure units.

There are four challenges to this goal.
First, it is hard to represent failure units uniformly for further analysis because 1) failure units can contain different numbers of metrics, and 2) feature engineering for various metrics is hard.
Second, due to the complex dependencies in an online service system, the faulty failure units can cause other metrics on the same or other components abnormal, and it is challenging to model the complex and various failure propagation. 
Third, since it is unlikely that all failure units have been faulty, it is essential but challenging to generalize to \textit{previously unseen failures} (no failures of the same kinds have occurred at the same locations).
Fourth, both local (interpreting individual cases) and global (interpreting general model decisions) interpretability~\cite{molnar2020interpretable} are important for engineers to trust the localization results,  since they provide interpretation from different perspectives.

This paper proposes an actionable and interpretable fault localization approach, \ours{}\footnote{A French phrase translating literally to ``already seen''}, for recurring failures in online service systems.
\textit{Recurring failures} are repeated failures of the same kinds at different locations (e.g., high response time caused by different inefficient SQL queries). 
Failures may recur due to misunderstanding of root causes, delayed fix deployment or emergent behaviors caused by high utilization \cite{bodik2010fingerprinting}.
Fault localization for recurring failures is important due to its large prevalence (e.g., 74.38\% in \ccb{}) in practice (see \cref{sec:recurring-failure-motivation}).
For recurring failures in a specific online service system, its engineers can summarize the indicative metrics on each class of components to recognize the underlying failure types and direct their mitigation action, according to their domain knowledge and diagnosing experience.
Based on these groups of indicative metrics, we define the candidate failure units for recurring failures in the system.
Furthermore, to represent the complex dependencies in the system, we connect the failure units that have dependencies between each other into a \textit{failure dependency graph (FDG)} (see \cref{sec:preliminary}).
Note that a system’s FDG is evolving due to deployment and software changes, and engineers can also add new failure units.

When a failure occurs, the monitoring system raises alerts and triggers \ours{}, which takes the latest FDG and the metric values as inputs and recommends suspicious failure units from the candidate failure units on the FDG to engineers (see later in \cref{fig:workflow}).
For challenge 1, \ours{} employs gated recurrent unit-based~\cite{cho2014learning} feature extractors to represent each failure unit as a fixed-width vector (unit-level feature) regardless of its metrics.
For challenge 2, we apply graph attention networks~\cite{velickovic2018graph} on the FDG to consider both the dependencies and the unit-level features.
For challenge 3, our model learns from the historical failures of the same kind but at \textit{any locations} to recommend a faulty failure unit by the metrics values and relative structure on the FDG only.
For challenge 4, we provide local interpretation by finding the representative historical failures from which the trained model probably learns to make the recommendations for a failure.
We also globally interpret the trained model as human-readable rules (e.g., see later in \cref{fig:global-interpretation-example}) by mimicking it with decision trees \cite{zhou2021machine}.
After engineers get the faulty failure units, they can timely recognize the underlying failure types and take mitigation action to ensure the quality of software services.
Finally, the ground-truth failure units manually confirmed by engineers will be saved for future retraining.
Unlike existing works \cite{brandon2020graphbased,ma2020diagnosing,bodik2010fingerprinting}, we do not aim to find similar historical failures and adopt their ground truths because such methods cannot localize \textit{previously unseen} failures.

We extensively validated \ours{} with four datasets, three of which are based on real-world systems, containing \num{502} injected failures of \num{18} types and \num{99} real-world failures\footnote{The datasets and implementation can be found in our replication package~\cite{2022dejavu}.}.
The results show that \ours{} is effective in localizing faulty failure units.
Specifically speaking, the average rank of the ground truths achieves 1.66$\sim$5.03 and outperforms baselines by 54.52\%$\sim$97.92\%.
The results also show that the main modules, such as feature extractors and GAT-based aggregation, indeed contribute.
Particularly, the results on production systems and real-world failures demonstrate practical performance.
\ours{} is efficient as it costs tens of minutes to train the localization model and less than one second to localize for a failure.
Moreover, \ours{} can achieve similar performance on previously unseen failures compared with seen failures.
Finally, we demonstrate the effectiveness of our interpretation techniques.

The contributions of this paper are summarized as follows:
\begin{itemize}[leftmargin=1em]
    \item For the first time, we propose an actionable and interpretable fault localization approach, \ours{}, for recurring failures in online service systems. \ours{} offline trains a localization model using historical failures in a given online service system and online recommends and interprets faulty failure units. 
    
    \item We propose a novel localization model and two interpretation methods addressing all four challenges.
    
    \item We conduct extensive experiments on \num{601} failures from four systems, including \num{99} real-world failures and three real-world systems. The results show \ours{}'s effectiveness, efficiency, generalizability, and interpretability.
    We also share lessons learned from our industrial experience.
\end{itemize}
\section{Background}
\label{sec:background}

\subsection{Recurring Failures}
\label{sec:recurring-failure-motivation}

\begin{figure}[htb]
\captionsetup{skip=0pt}
\centering
\includegraphics[width=0.9\columnwidth]{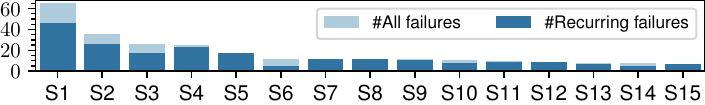}
\caption{The number of recurring failures at \ccb{}.}
\Description{In the systems at \ccb{}, most failures are recurring failures.}
\label{fig:ccb-recurring-failures-count}   
\end{figure}

We investigated the failures tickets at \ccb{}, a large commercial bank with over a hundred million users, to motivate our work on recurring failures.
The banking information system at \ccb{} contains over \num{300} applications, each of which runs on dozens of servers and contains many components such as databases, web servers, and load balancers.
We collected \num{576} failure tickets from this system, spanning 12 months, which are triggered by application SLO (service level objective) violation and contain a detailed diagnosis and mitigation process.

We categorize the root causes of these failures into recurring and non-recurring categories by whether they can be repetitive with historical failures (maybe at different locations).
The recurring categories are mainly external, hardware, or middleware reasons, such as \textit{bad requests}, \textit{unavailable third-party services}, \textit{failed disks}, \textit{slow SQL queries}, and \textit{missing database indices}.
The non-recurring categories are mainly logical reasons, such as \textit{code defects}, \textit{design flaws}, and \textit{data inconsistency} (e.g., incorrect modification on a manually maintained special account list).

In \cref{fig:ccb-recurring-failures-count}, we present the number of all/recurring failures in the applications with the most failure tickets.
On all applications except S6, most failures (at least 2/3) are recurring.
In total, there are 74.38\% recurring failures. 
Lots of failures (39.28\%) are recurring due to unavailable third-party services.
There are also existing studies reporting the large prevalence of recurring failures.
For example, \citeauthor{dogga2022revelio} find that recurring categories of root causes cause 94\% of the failures at a major SaaS company~\cite{dogga2022revelio}, and \citeauthor{lee2000diagnosing} find that 70\% of the failures they studied are recurring~\cite{lee2000diagnosing}.

In summary, diagnosing recurring failures is important due to the large prevalence.
Moreover, the repetitive nature motivates us to diagnose recurring failures by learning from historical failures.

\subsection{Defining Failure Units and FDGs}
\label{sec:preliminary}

\begin{figure}[htb]
    \captionsetup{skip=0pt}
    \centering
    \includegraphics[width=0.95\columnwidth]{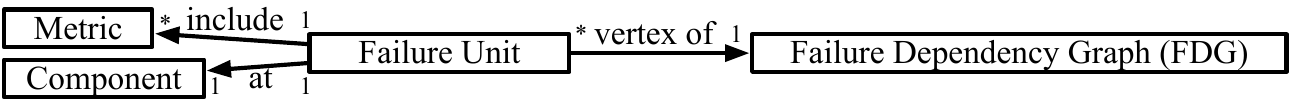}
    \caption{The relationship of the basic concepts}
    \label{fig:basic-concepts}
    \Description{The relationship of the basic concepts}
\end{figure}

\begin{figure*}[!tbh]
\captionsetup{skip=0pt}
\includegraphics[width=1.7\columnwidth]{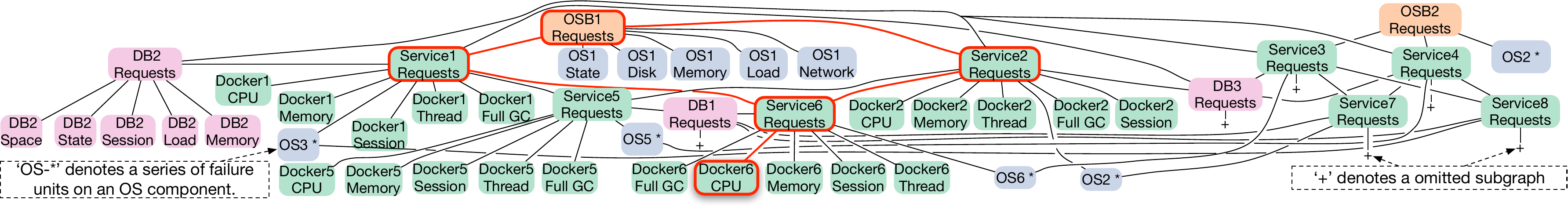}
\caption{The FDG of \sys{A}. There are 102 vertices and 109 edges in total, some of which are omitted due to the space limit.}
\label{fig:sys-a-FDG}
\Description{A failure dependency graph of system A}
\end{figure*}

\begingroup
\setlength{\tabcolsep}{1pt} %
\renewcommand{\arraystretch}{1} %
\begin{table}[htb]
\caption{Component classes (CC) and metric groups in \sys{A}
}
\footnotesize
\begin{tabularx}{\columnwidth}{ccX}
\thickhline
    \textbf{CC} & \textbf{Group} & \textbf{Metrics in the metric group} \\ \thickhline
    \begin{tabular}[c]{@{}l@{}}DB, OSB, \\Service\end{tabular} & Requests & \#requests, average response time, success rate, process time \\ \hline
    DB &  Load & ACS, AIOS, AWS, CPUUsedPct, CPUFreePct, CPUPused, \{Call, DFParaWrite, Exec, LFParaWrite, LFSync, Login, TPS, SctRead\}PerSec, Logic/Physical ReadPerSec, PGAUsedPct, SEQUsedPct, SessConnect, TbsUsedPct, UndoTbsPct, UserCommit \\ \hline
    DB &  Memory & MEMTotal, MEMUsed, MEMUsedPct, MEMRealUtil \\ \hline
    DB &  Session & ProcUsedPct, ProcUserUsedPct, Sess\{Active, Connect, UsedTemp, UsedUndo, Pct\} \\ \hline
    DB &  State & DbTime, Hang, OnOffState, RowLock, tnspingResultTime\\ \hline
    DB &  Space & AsmFreeTb, DbFileUsedPct, NewTbs FreeGb/UsedPct, PGAUsedTotal, RedoPerSec, TbsFreeGb, TbsUsedPct, TempTbsPct, Total/Used TbsSize \\ \hline
    Docker &  CPU & containerCpuUsed\\ \hline
    Docker &  Memory & containerMemUsed\\ \hline
    Docker &  Session & containerSessionUsed\\ \hline
    Docker &  Thread & Thread\{Idle, Running, Total, UsedPct\}\\ \hline
    Docker &  Full GC & containerFgc, containerFgct\\ \hline
    OS &  State & AgentPing, ICMPPing \\ \hline
    OS &  Disk & Disk\{AvgquSz, Await, IoUtil, RdIos, RdKbs, Svctm, WrIos, WrKbs\}, FS\{MaxAvail, MaxUtil, TotalSpace, UsedPct, UsedSpace\}, Free/Total DiskSpace, Free/Total Inodes, Used DiskSpace/Inodes, Used DiskSpace/Inodes Pct \\ \hline
    OS &  Load & BuffersUsed, CPU\{Frequency, IdlePct, IowaitTime, KernelNumber, Number, Pused, SystemTime, UserTime, UtilPc, UtilPct\}, NumOfProcesses, NumOfRunningProcesses, ProcessorLoad 1/5/15 Min, System Block/Wait Queue Length, ZombieProcess \\ \hline
    OS &  Memory & BuffersUsed, CacheUsed, MemoryAvailable, MemoryAvailablePct, MemoryFree, MemoryTotal, MemoryUsed, MemoryUsedPct, PagePi, PagePo, SharedMemory, SwapUsedPct \\ \hline
    OS &  Network & Incoming/Outgoing NetworkTraffic, Received/Sent ErrorsPackets/Packets/Queue/Total, ssTotal \\
\thickhline 
\multicolumn{3}{l}{* \#components in OSB (Oracle Service Bus)/Service/Docker/DB/OS: 2/8/8/3/6.}
\end{tabularx}
\label{tbl:failure-class-in-sys-a}
\end{table}
\endgroup

This section describes how \ours{} defines failure units and FDGs to enable actionable fault localization.
As introduced in \cref{sec:introduction}, the experienced engineers of an online service system can define the candidate failure units by summarizing the indicative metric groups on different component classes.
For example, in \cref{tbl:failure-class-in-sys-a}, we present the component classes and metric groups for \ds{A} (see \cref{sec:experiment-setup}).
\textit{Each group of such indicative metrics at a corresponding component is a candidate \textit{failure unit} for recurring failures in the system}.
For example, the metrics in \textit{Requests} (i.e., \textit{\#requests}/\textit{average response time}/...) at DB1 form a failure unit, \textit{DB1 Requests}.
In this paper, we focus on metrics, as logs are of huge volume and various types and traces contain little information on low-level performance issues.
For convenience, we name the failure units defined by the same group of metrics on the same component class, which contain the same metrics at different locations, as a \textit{failure class}.

To model the complex and various failure propagation, we represent the dependencies in an online service system with an \textit{FDG}.
\begin{definition}
An FDG (failure dependency graph) is an undirected graph, $G=(V, E)$. 
$V$ is the set of all the defined candidate failure units of an online service system.
An edge $(v_i, v_j)$ exists in the edge set, $E$, \textit{iff.} the failure unit $v_i$ depends on $v_j$ or vice versa.
\end{definition}
For example, the FDG of \sys{A} (\cref{sec:experiment-setup}) is shown in \cref{fig:sys-a-FDG}, which is complex considering that \sys{A} comprises only 27 components (see \cref{tbl:failure-class-in-sys-a}).
In \cref{fig:sys-a-FDG}, a failure caused by CPU exhaustion on \textit{Docker6} propagates to \textit{Service6 Requests}, since \textit{Service6} is deployed on \textit{Docker6}.
It further propagates to \textit{Service1 Requests}, \textit{Service2 Requests}, and \textit{OSB1 Requests}, since \textit{Service1} and \textit{Service2} rely on \textit{Service6} and \textit{OSB1} relies on these two services.

\begin{figure}[htb]
    \captionsetup{skip=0pt}
    \centering
    \includegraphics[width=\columnwidth]{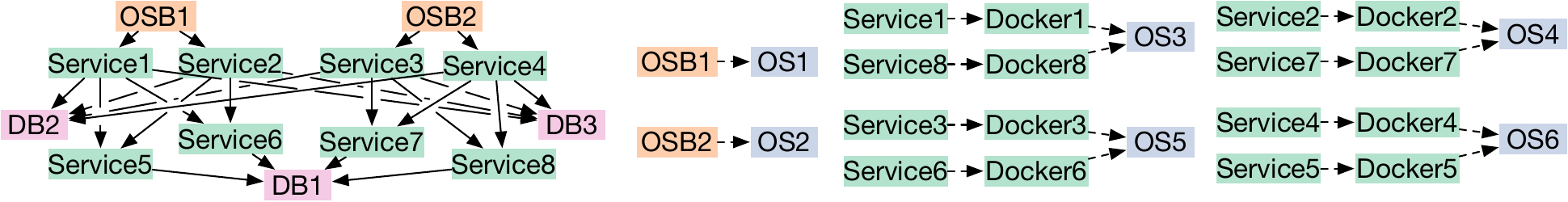}
    \caption{The call (solid) and deployment (dashed) component relationships of \sys{A}}
    \label{fig:sys-a-component-relationship}
    \Description{The relationships among components are used to define an FDG but they are much simpler than the FDG.}
\end{figure}

We construct FDGs automatically, which is necessary because FDGs are complex and dynamic (e.g., in microservice systems, pods are dynamically created and deleted), based on the call and deployment component relationships.
For example, in \sys{A}, we collect the call relationships among OSBs, services, and databases through tracing tools (the leftmost graph in \cref{fig:sys-a-component-relationship}) and collect deployment relationships among services, containers, and servers (the others in \cref{fig:sys-a-component-relationship}) through configuration management database.
By combining the component relationships and failure units, we automatically construct an FDG.
For example, as \textit{Service1} is deployed on \textit{Docker1} (see \cref{fig:sys-a-component-relationship}), \textit{Service1 Requests} is connected to \textit{Docker1 CPU/Memory/...} (see \cref{fig:sys-a-FDG}).
Besides, engineers can manually add or remove edges on the FDG according to their domain knowledge and existing diagnosis knowledge.
For example, when two services depend on a stateful third-party service, which cannot be captured by tracing, engineers can manually connect them on the FDGs.

In this paper, we group metrics roughly by their general categories (e.g., CPU-related or network-related) for most datasets and include no prior knowledge in the construction of FDGs.
In practice, engineers can carefully tune the specifications of failure units and FDGs to make the localization results more helpful, which is out of the topic of this work.

\subsection{Industrial Localization Practice}
\label{sec:industrial-localization-practice}
The current industrial fault localization is largely manual.
We have analyzed over 20,000 failure tickets in a large commercial bank spanning two years. 
We found that the average diagnosis time is 28.98 minutes, and the 25/50/75 percentile is 4.27/9.65/25.81 minutes.
Notably, the diagnosis time could be extremely long when the failure is triaged among many teams.

Furthermore, the average time cost for engineers to localize the fault locations and types (i.e., failure units) is 9.2 minutes. 
The 25/50/75 percentile is 2.8/5.0/10.0 minutes.
As a result, localizing faulty failure units is promising to save much time for engineers.
Here are some examples:
\begin{itemize}[leftmargin=1em]
 \item At 21:23, the on-call engineer (OCE) was alerted that a server had suffered from a high I/O delay. At 21:26, the OCE confirmed a failed disk caused the failure by checking logs.
 \item At 17:27, the OCE was alerted that the success rate of service A was low. At 17:32, the OCE confirmed that a problematic third-party service caused the failure by checking the success rate under different conditions (e.g., user request type, client version, and client location).
 \item At 09:45, the OCE was alerted of the low success rate of service B. The OCE checked the owned systems and found no problems, and thus, called the OCEs of the related third-party system for cooperation. At 10:30, the OCEs of a third-party system confirmed the fault lay in their system.
\end{itemize}

\subsection{Problem Statement}
\label{sec:problem-statement}

In this work, we target recommending the faulty failure unit given the latest FDG and the corresponding metric values when a failure occurs.
The faulty failure unit pinpoints where the failure occurs (the component) and which kind of failure occurs (indicated by the metrics), and thus, helps engineers take the right mitigation actions rapidly.
Finding the most helpful specifications of failure units or FDGs and failure discovery are out of the scope of this paper.
\section{Design of \ours{} model}
\label{sec:RCL-model-design}

\begin{figure}[htb]
\includegraphics[width=\columnwidth]{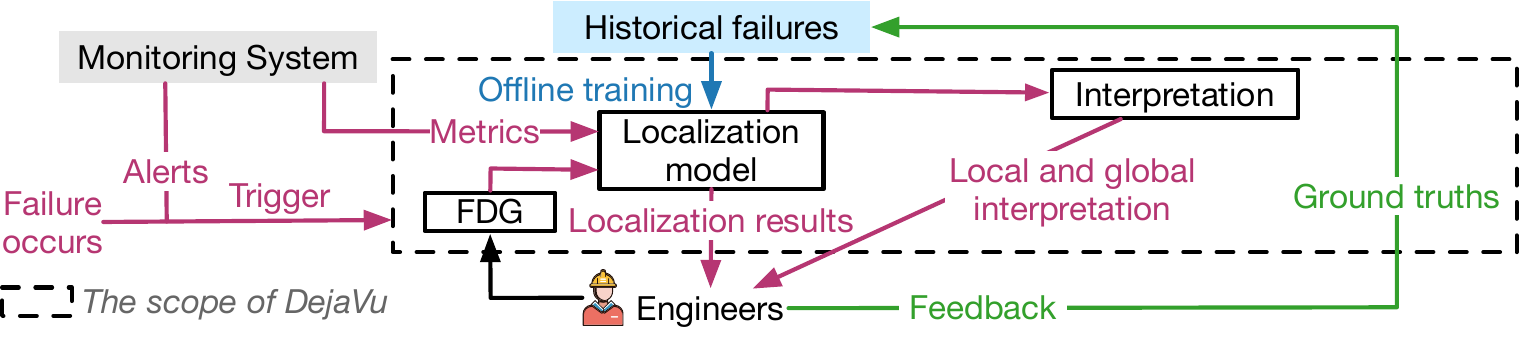}
\caption{Workflow of \ours{}}
\label{fig:workflow}
\Description{Workflow of \ours{}}
\end{figure}

In this section, we introduce the design of \ours{} model.
As shown in \cref{fig:workflow}, we train a model offline with historical failures and FDGs of a given online service system.
The trained model recommends faulty failure units online when a failure occurs, given the metric values and the latest FDG.

\subsection{Overview}
\label{sec:model-overview}

\begin{figure}[htb]
    \captionsetup{skip=0pt}
    \includegraphics[width=\columnwidth]{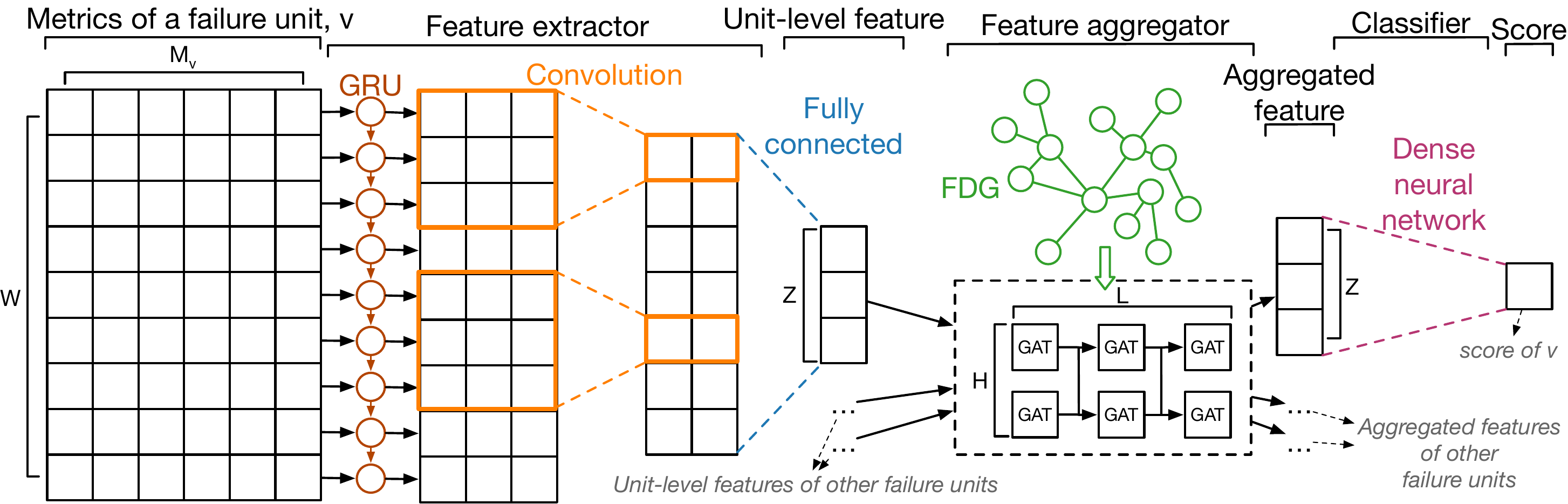}
    \caption{\ours{} model architecture}
    \label{fig:model-architecture}
    \Description{The model architecture}
\end{figure}

As shown in \cref{fig:model-architecture}, the \ours{} model takes the metric values and the FDG around the failure time as inputs and outputs a suspicious score for each failure unit with a binary classifier.
Note that different failures could have different corresponding FDGs.

More specifically, first, to solve the challenge of representing failure units uniformly, we introduce a feature extractor module that can grasp temporal information of metrics and correlations among metrics (\cref{sec:feature-extractor}).
A feature extractor is trained to map the metrics of any failure unit of the same failure class into a fixed-width vector (unit-level feature).
We train a failure extractor for each failure class, since failure units of different failure classes contain different metrics.
Second, to enable modeling failure propagation, we employ a feature aggregator to encode the structural information on the FDG into aggregated features (\cref{sec:feature-aggregator}).
The feature aggregator utilizes attention mechanism to pay more attention on related failure units.
Finally, we introduce a classifier to score each failure unit based on its aggregated feature (\cref{sec:classifier}).
For generalizability, the failure units in the same failure class share a feature extractor, and all failure units share a feature aggregator and classifier.
For different failure classes, their feature extractors are all the same except the input size.
In this way, the score of a failure unit (e.g., \textit{Docker6 CPU} in \cref{fig:sys-a-FDG}) is majorly determined by its metric values and the metric values of its related failure units (e.g., \textit{Service6 Requests}, \textit{Service2 Requests}, \textit{OSB1 Requests}) rather than its location.

Our model is trained by minimizing a revised loss function (\cref{sec:loss-function}).
Furthermore, we employ class balancing for training, since the frequency of each failure class in training data varies (\cref{sec:class-balancing}).

\subsection{Feature Extractor}
\label{sec:feature-extractor}
Each metric of a failure unit has temporal information, and there are correlations among the metrics.
To better extract meaningful features, we use a three-stage feature extraction method.
Specifically speaking, the first stage learns temporal information, while the latter two learn higher-level features.

In the first stage, the temporal feature of a failure unit is extracted using gated recurrent unit (GRU)\cite{cho2014learning} recurrent neural networks.
GRU is a recurrent neural network with a gating mechanism but has simpler architecture and fewer parameters than LSTM.
The input metrics of a failure unit are encoded as a numerical matrix of dimension $W \times M_{v}$, where $W$ is the length of time window we slice at the failure time and $M_{v}$ is the number of metrics of the failure unit $v$.
In this paper, we empirically set $W$ as 20 minutes to capture the near history of the metrics.

In the second stage, we apply 1-D convolution neural network (1-D CNN)~\cite{goodfellow2016deep} and GELU (Gaussian error linear unit) activation function \cite{hendrycks2020gaussian} on the temporal feature matrix obtained at the first stage.
With the application of 1-D CNN, we get multiple feature maps extracting the correlations among different time points and metrics. 
GELU provides nonlinearity by preserving linearity in the positive activations and suppressing the negative activations while relieving vanishing gradients problem \cite{hendrycks2020gaussian}.

In the third stage, a fully connected layer is applied to learn the relations among different feature maps obtained in the second stage and output a numeric \textit{unit-level feature} vector of dimension $Z$.

\subsection{Feature Aggregator}
\label{sec:feature-aggregator}
The second module, feature aggregator, is responsible for aggregating the unit-level features of related failure units of a failure unit $v$ and the structure information into one \textit{aggregated feature}, $\boldsymbol{\hat{f}}^{(v)}$, based on the FDG, $G$.
For generalizability, the feature aggregator should be structure-independent.
Thus, graph convolutional network (GCN) \cite{kipf2017semisupervised} is unsuitable.
Since the relationship between any pair of failure units varies, it is also inappropriate to simply average features of connected failure units together as GraphSAGE \cite{hamilton2017inductive} does.
Therefore, for each failure unit, we dynamically calculate the aggregation weights of its related failure units based on their unit-level features through graph attention networks (GAT)\cite{velickovic2018graph}.

However, a single GAT aggregates features of a failure unit's neighbors only, while failures propagate along the FDG for more than one hop.
Therefore, we stack multiple GAT together sequentially, and the outputs of the last GAT are taken as the input of the next GAT.
In this way, we aggregate features of farther related failure units and thus, model multi-hop failure propagation.
Though multiple-layer graph neural network could suffer from over-smoothing, we apply residual connections to relieve the problem~\cite{li2019deepgcns}.
On the other hand, we also introduce multi-head attention to improve the capacity and stabilize the training process \cite{velickovic2018graph}.
Specifically speaking, we apply multiple GATs in parallel and concatenate their outputs together.
We denote the number of sequentially stacked GATs as $L$ and the number of heads as $H$.
We set $H=4$ and $L=8$ by default and discuss their impact in \cref{sec:configuration-impact}.
For generalizability, the feature aggregator is shared by all failure units regardless of their classes.
In summary, the feature aggregator can be formalized as follows.
\begin{equation}
\begin{aligned} 
    \textstyle &\{\boldsymbol{\hat{f}}^{(l+1, v)}|v {\in} V\}{=}\underbrace{[\text{GAT}(\{\boldsymbol{\hat{f}}^{(l, v)}|v {\in} V\});...;\text{GAT}(\{\boldsymbol{\hat{f}}^{(l, v)}|v {\in} V\})]}_{H\>\text{GATs in total}}
\end{aligned}\nonumber
\end{equation}
where $l \in \{0, 1, ..., L-1\}$, $\boldsymbol{\hat{f}}^{(0, v)}{:=}\boldsymbol{{f}}^{(v)}$, $\boldsymbol{\hat{f}}^{(v)}{:=}\boldsymbol{\hat{f}}^{(L, v)}$, and $GAT(\cdot)$ calculates aggregated features for all failure units in the FDG.

\subsection{Classifier}
\label{sec:classifier}
The final module, classifier, assigns a suspicious score, $s(v)$, for each failure unit given its aggregated feature, $\boldsymbol{\hat{f}}^{(v)}$.
Specifically speaking, we simply employ a two-layer dense neural network since it is already enough to achieve satisfactory performance.
To restrict the output value in $[0, 1]$, we use sigmoid function $\sigma$ as the output activation.
In summary, the output suspicious score for a failure unit $v$ (denoted as $s(v)$) can be formulated as follows,
\begin{equation}
    s(v)=(\sigma\circ\text{Dense}\circ\text{GELU}\circ\text{Dense})(\boldsymbol{\hat{f}}^{(v)})
\label{eqn:classifier}
\end{equation}
We do not need thresholds on the suspicious scores to obtain binary classification for failure units.
Instead, we only rank all failure units by their suspicious scores in descending order, based on which engineers can check the suspicious failure units one by one.

\subsection{Loss Function}
\label{sec:loss-function}

For each failure unit $v$ of a failure $T$, we expect the suspicious score $s_{T}(v)\in[0, 1]$ to be as close as possible to its ground truth label $r_{T}(v) \in \{0, 1\}$.
To measure the difference between $s_{T}(v)$ and $r_{T}(v)$, we use the widely-used binary cross-entropy \cite{goodfellow2016deep}:
$$
BCE(r_{T}(v){,}s_{T}(v))=r_{T}(v)\cdot\log\big(s_{T}(v)\big)+(1{-}r_{T}(v))\cdot\log\big(1{-}s_{T}(v)\big) 
$$
Since most failure units are not faulty, their losses are dominant compared with the faulty ones.
As a result, the model could fall back to score every failure unit equally.
To solve this problem, we assign different weights to faulty and normal failure units to make the weight of faulty failure units equal to the sum of all normal failure units.
Specifically speaking, the weights of normal and faulty failure units are $1$ and $N$ (the number of failure units), respectively.
In summary,
\begin{equation}
    \textstyle \mathcal{L}_{s}=\frac{1}{N_H}\sum_{T \in \{T_1, \cdots, T_{N_{H}}\}}[\frac{\sum_{v\in V}\boldsymbol{w}^{(T)}_{v}BCE\big(r_{T}(v), s_{T}(v)\big)}{\sum_{v\in V}\boldsymbol{w}^{(T)}_{v}}]
\end{equation}
where $V$ is the set of failure units, $\boldsymbol{w}^{(T)}_{v}=r_{T}(v)\cdot |V| + (1 - r_{T}(v))\cdot 1$, and $ \{T_1, \cdots, T_{N_{H}}\}$ is the set of training failures.
By minimizing $\mathcal{L}_{s}$ on all historical failures, we train the \ours{} model.

\subsection{Class Balancing}
\label{sec:class-balancing}
In practice, the number of failures in different classes varies.
Thus, to prevent the \ours{} model from neglecting the minority classes in the training process, we upsample historical failures of such classes.
Suppose there are $C$ failure classes and the number of failures of the $i$-th class is $N_H^{(i)}$, we sample a failure with probability $1 / (C \cdot N_H^{(i)})$ for each training step.
In this way, the failures of each class are sampled with the same probability (i.e., $1 / C$).

\section{Interpretation Methods}
\label{sec:interpretation-methods}

\subsection{Global Interpretation}
\label{sec:global-interpretation}

\begin{figure}[htb]
\captionsetup{skip=0pt}
\centering
\includegraphics[width=0.9\columnwidth]{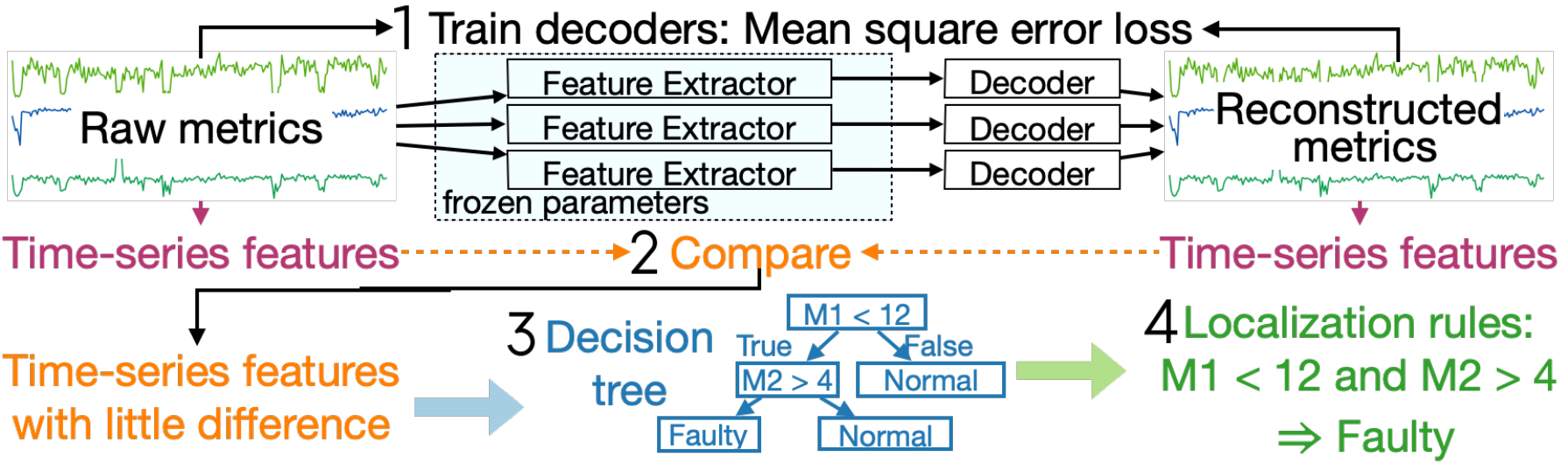}
\caption{Global interpretation}    
\label{fig:global-interpretation}
\Description{Illustration of our global explanation method}
\end{figure}

Our core idea is to use simpler but interpretable models to approximate (rather than outperform) the sophisticated black-box DejaVu models as accurately as possible.
As shown in \cref{fig:global-interpretation}, we use decision trees (DTs) as the surrogate models.
The inputs for DTs are selected interpretable time-series features of failure units' metrics, and targets are the suspicious scores from \ours{}.
We do not use existing deep-learning model interpretation methods for two reasons.
On the one hand, though many interpretation methods~\cite{zeiler2014visualizing,bau2017network,dicicco2019interpreting} are targeted at interpreting deep-learning models by understanding their inner mechanism, engineers are mainly concerned about the relationship between raw data and root causes~\cite{meng2020interpreting}.
On the other hand, the inputs and targets of our problem are non-standard and require extra effort.

First, we select 57 easy-to-understand time-series feature extractors (e.g., \texttt{range\_count} and \texttt{variance}) from \texttt{tsfresh}\cite{2021tsfresh}.
They would extract \num{191} time-series features for each metric (see our replication package~\cite{2022dejavu}), each of which represents an understandable property of a metric, such as level and variance.

However, the number of time-series features for a failure unit (i.e., $\#\text{metrics}\times 191$) can be extremely large (e.g., thousands), while the feature dimension of \ours{} (i.e., $Z$) is much lower (e.g., three).
It indicates that \ours{} cannot consider most of these time-series features.
To remove the useless and noisy time-series features, for each failure class, we first train a decoder to reconstruct metrics from the unit-level features given by \ours{} and then compare the time-series features extracted from the original and reconstructed metrics.
A decoder comprises a one-dimensional transpose convolution (1-D DeConv) layer, which can be treated as a transpose operation of Conv1D \cite{dumoulin2016guide}, and a fully-connected layer, and is trained by minimizing the mean square error between the original and reconstructed metrics.
If a time-series feature differs a lot in the original and reconstructed metrics, it is not reserved by the \ours{} model and should be removed.

The training sample for decision trees of a failure unit comprises the selected time-series features of its metrics (as input) and the categorized suspicious score given by the \ours{} model (as target).
When interpreting a \ours{} model, we are not concerned about the knowledge on which it is unconfident.
Thus, we categorize the suspicious scores (i.e., $s(v)$ in \cref{eqn:classifier}) into three different categories.
Specially speaking, if $s(v)>0.9$, then failure unit $v$ is classified as a \textit{faulty} failure unit; if $s(v)<0.1$, then failure unit $v$ is a \textit{normal} failure unit; otherwise, the \ours{} model is \textit{uncertain}.
With the training samples, we train decision trees which mimic the \ours{} model (part 3 in \cref{fig:global-interpretation}) and classify failure units as ``faulty'' or ``normal.''
Since failure units in different failure classes contain different metrics, we train a decision tree for each failure class.

On the trained decision trees, we focus on the decision paths leading to only faulty or normal failure units (part 4 in \cref{fig:global-interpretation}).
These paths are the diagnosis rules that the \ours{} model learns from historical failures.
The rules are used to help engineers understand and trust the \ours{} model rather than replace it.
On the one hand, the performance of the decision trees is still worse than \ours{}. 
On the other hand, the interpretations provided by the decision trees can also make mistakes~\cite{meng2020interpreting}.
As a result, engineers are supposed to pay attention to the qualitative meaning of split conditions (e.g., the success rate of an external service is very low) rather than the specific thresholds.

\subsection{Local Interpretation}
\label{sec:local-interpretation}

Since the \ours{} model is trained with historical failures, it is straightforward to interpret how it diagnoses a given failure by figuring out from which historical failures it learns to recommend the faulty failure units.
To achieve this goal, we compare the incoming failure with each historical failure based on the aggregated features extracted by the trained \ours{} model.
The comparison is conducted on each pair of failure classes instead of failure units, since two similar failures can occur at different localization (thus have different faulty failure units).
Due to the space limit, we omit the details, which can be found in our code~\cite{2022dejavu}.

\section{Experiments}
\label{sec:experiments}
In this section, we aim to address the following research questions.
\begin{itemize}[leftmargin=1em]
    \begin{minipage}{\columnwidth} \item \textbf{RQ1}: How does \ours{} perform in fault localization? \end{minipage}
    \begin{minipage}{\columnwidth} \item \textbf{RQ2}: How does \ours{} perform in various situations? \end{minipage}
    \begin{minipage}{\columnwidth} \item \textbf{RQ3}: How does \ours{} perform in terms of \textit{efficiency}? \end{minipage}
    \begin{minipage}{\columnwidth} \item \textbf{RQ4}: How does \ours{} perform on previously unseen failures? \end{minipage}
    \begin{minipage}{\columnwidth} \item \textbf{RQ5}: How do our interpretation methods perform? \end{minipage}
\end{itemize}

\subsection{Experiment Setup}
\label{sec:experiment-setup}

\subsubsection{Study Data}
\label{sec:dataset}

\begingroup
\setlength{\tabcolsep}{4pt} %
\renewcommand{\arraystretch}{1} %
\begin{table*}[t]
\captionsetup{skip=0pt}
\caption{Dataset summary}
\footnotesize
\begin{tabular}{ccccccc}
\thickhline
    Dataset & \#Failures & \#Metrics & \#Failure units & \#Failure Classes & System & Failure Source\\
    \hline
    \ds{A} & \num{188} & \num{710} & \num{102}& \num{18} & A production microservice system of a major ISP (\sys{A}) & Injected by the engineers\\
    \ds{B} & \num{158} & \num{2419} & \num{189} & \num{20} & A production serivce-oriented system of a commercial bank (\sys{B}) & Injected by the engineers\\
    \ds{C} & \num{99} & \num{2594} & \num{41} & \num{41} & A production Oracle database system of a commercial bank (\sys{C}) & Real-world failures\\
    \ds{D} & \num{156} & \num{5724} & \num{1044} & \num{23} & A open-source microservice benchmark system, Train-Ticket~\cite{zhou2018fault} (\sys{D}) & Injected by us \\
    \thickhline
\end{tabular}
\label{tbl:dataset}    
\end{table*}
\endgroup

In the study, we use four datasets (\ds{A}, \ds{B}, \ds{C}, and \ds{D}), containing \num{601} failures in total.
In particular, \num{16} failures have multiple root causes and thus multiple faulty failure units.
\ds{A}, \ds{B}, and \ds{C} are from production systems, and \ds{D} is based on an open-source benchmark system.
The statistics of them are shown in \cref{tbl:dataset}.
In all datasets, the FDGs are automatically constructed without any manual modification following the process in \cref{sec:background}.
To obtain failures for training, we split each dataset into a training set (40\%), a validation set (20\%), and a testing set (40\%).
All experiment results are calculated on the testing sets only.

The ground truths of \ds{A}, \ds{B}, and \ds{D} are determined by the locations and types of failure injection.
In \ds{A} and \ds{B}, their engineers injected ten types of failures: 
1) CPU exhaustion on containers, physical servers (only \ds{B}), or middlewares (only \ds{B});
2) packet loss or delay on physical servers;
3) database session limit (only \ds{A}) or halt (only \ds{A});
4) low free memory for JVM/Tomcat (only \ds{B});
5) disk I/O exhaustion (only \ds{B}).
For \ds{D}, we deploy Train-Ticket~\cite{zhou2018fault} on a 4-node Kubernetes cluster.
Train-Ticket is one of the largest open-source microservice benchmarks, containing 64 services.
Following existing works~\cite{yu2021microrank,liu2020unsupervised}, we performed eight types of failure injection at random locations with ChaosMesh~\cite{2022chaosmesh}: CPU/memory stress on pods/nodes, pod failure, and packet corrupt/loss/delay on pods.
Details of the deployment and failure injection for \ds{D} can be found in our replication package~\cite{2022dejavu}.

For \ds{C}, we collected \num{99} real-world failures on \sys{C}, spanning several months, and we labeled their ground-truth failure units with the engineers.
Different from the other three, \sys{C} is not a typical online service system and has only one component.
Thus, each failure class in \ds{C} contains only one failure unit.
The definition of the failure units in \ds{C} are confirmed by the experienced engineers.
In the FDG of \sys{C}, all failure units are only connected to a virtual vertex (there is no metric in it), as we do not assume any prior diagnosis knowledge on the causal dependency among the metrics.

\subsubsection{Baselines}
\label{sec:baselines}
We compare \ours{} with the following state-of-the-art baseline methods: 
\begin{itemize}[leftmargin=1em]
    \item JSS'20 \cite{brandon2020graphbased} represents failures as graphs and finds similar diagnosed failures by graph similarity for each incoming failure.
    \item iSQUAD \cite{ma2020diagnosing} clusters historical failures, labels the root causes of each cluster, and assigns each incoming failure to a cluster.
    \item Traditional machine learning (ML) models, including Decision Tree (DT), Gradient Boosting (GB), Random Forest (RF), and SVM. For each failure class, we train a model which takes the time-series features (see \cref{sec:global-interpretation}) of a failure unit and determines whether it is faulty.
    \item Random walk (RW) is widely used in unsupervised heuristic fault localization. As existing random walk-based works \cite{kim2013root,wu2021microdiag} cannot take various numbers of metrics as input, we compare with two variants:
        \begin{itemize}[leftmargin=1em]
        \item RandomWalk@Metric. Following existing works, we first construct a causal graph, calculate transition probabilities based on metric correlations, and obtain the score of each metric with personalized PageRank. Then, we obtain the score of each failure unit by summing up the scores of its metrics.
        \item RandomWalk@FI. We score failure units (FI) directly by applying random walk on the FDGs and using the average correlation of all pairs of metrics of two failure units as transition probabilities.
        \end{itemize} 
\end{itemize}
The aggregation methods for RandomWalk@Metric (sum) and RandomWalk@FI (average) are selected so as to maximize the overall performance.
Some methods~\cite{zhou2019latent,gan2021sagea,li2021practical,liu2020microhecl} are not applicable because they localize faulty services only.

\subsubsection{Implementation}
\label{sec:implementation}
We implement \ours{} and all baseline methods based on \texttt{PyTorch}\cite{2021pytorch} and \texttt{DGL}\cite{wang2019deep}.
We use Adam \cite{kingma2015adam} to train the \ours{} models and set the initial learning rate to \num{0.01} and weight decay to \num{0.01}.
To avoid gradient exploding, we clip the gradients of all parameters at \num{1}.
We train each model for \num{3000} epochs and set the batch size to \num{16}.
Our implementation, including the baselines, are public in our replication package~\cite{2022dejavu}.

\subsubsection{Evaluation Metrics}
\label{sec:evaluation-metrics}

Following related works \cite{zhou2019latent,liu2020microhecl}, we validate the effectiveness with top-$k$ accuracy ($A@k$) and mean average rank (MAR).
$A@k$ is the fraction of failures whose ground truths are included in the top-$k$ recommendations. 
It represents how many failures can be diagnosed by checking only the first top-$k$ recommendations, and larger is better.
MAR is the mean of the average suggested rank of all ground truths of each failure. 
It represents how many recommendations should be checked to diagnose a failure on average, and smaller is better.
Compared with $A@k$, MAR considers all recommended failure units but could be affected by extreme bad cases.
We also measure the efficiency with respect to localization and training time.

\subsection{RQ1: Effectiveness}
\label{sec:overall-performance}

\subsubsection{Overall Performance}
\begingroup
\setlength{\tabcolsep}{8pt} %
\renewcommand{\arraystretch}{1} %
\begin{table*}[htb]
\captionsetup{skip=0pt}
\caption{Fault localization results}
\footnotesize

\begin{tabular}{llllllllll}
\toprule
Dataset & Category & Method &          Effect Size &  p-value &     MAR &      A@1 &      A@2 &      A@3 &      A@5 \\
\midrule
\multirow{12}{*}{$\mathcal{A}$} & Supervised & DéjàVu &                    - &        - &    1.66 &  77.18\% &  90.38\% &  93.40\% &  96.28\% \\ \cline{2-3}
              & \multirow{2}{*}{Similar failure matching} & JSS'20 &   Huge       (40.75) &  5.4e-40 &   24.31 &  16.66\% &  25.64\% &  35.89\% &  47.44\% \\
              & & iSQUAD &   Huge       (10.90) &  2.3e-18 &   12.91 &  23.07\% &  30.77\% &  39.74\% &  46.16\% \\ \cline{2-3}
              & \multirow{4}{*}{Traditional ML} & Decision Tree &    Huge       (9.45) &  1.2e-40 &   18.05 &  53.20\% &  63.46\% &  65.77\% &  66.03\% \\
              & & Gradient Boosting &    Huge       (2.65) &  5.0e-14 &    3.82 &  63.62\% &  72.49\% &  76.18\% &  81.35\% \\
              & & Random Forest &    Medium     (0.51) &  3.3e-02 &    1.88 &  73.37\% &  89.15\% &  91.94\% &  95.71\% \\
              & & SVM &    Huge       (6.62) &  2.9e-32 &    7.66 &  31.54\% &  39.05\% &  44.53\% &  51.34\% \\ \cline{2-3}
              & \multirow{2}{*}{Unsupervised heuristic} & RandomWalk@Metric &   Huge       (12.23) &  4.3e-20 &   11.26 &   5.04\% &   8.89\% &  12.74\% &  26.73\% \\
              & & RandomWalk@FI &   Huge       (28.70) &  5.2e-34 &   33.64 &  16.51\% &  19.04\% &  20.32\% &  20.32\% \\ \cline{2-3}
              & \multirow{3}{*}{Ablation Study} & DéjàVu w/o GRU &    Large      (1.01) &  2.5e-04 &    3.25 &  74.36\% &  87.56\% &  91.67\% &  93.46\% \\
              & & DéjàVu w/o AGG &    Medium     (0.70) &  6.8e-03 &    2.32 &  76.80\% &  92.44\% &  94.49\% &  97.05\% \\
              & & DéjàVu w/o BAL &    Large      (1.04) &  1.8e-04 &    2.39 &  71.28\% &  85.00\% &  88.97\% &  92.44\% \\
\cline{1-10}
\multirow{12}{*}{$\mathcal{B}$} & Supervised & DéjàVu &                    - &        - &    5.03 &  66.21\% &  71.21\% &  75.61\% &  79.24\% \\ \cline{2-3}
              & \multirow{2}{*}{Similar failure matching} & JSS'20 &   Huge       (55.03) &  1.7e-17 &   47.92 &  15.15\% &  19.70\% &  24.24\% &  30.30\% \\
              & & iSQUAD &   Huge       (33.00) &  1.7e-15 &   30.75 &   7.58\% &  10.61\% &  15.15\% &  24.24\% \\ \cline{2-3}
              & \multirow{4}{*}{Traditional ML} & Decision Tree &   Huge       (16.29) &  1.3e-18 &   30.70 &  55.91\% &  60.30\% &  63.49\% &  65.00\% \\
              & & Gradient Boosting &    Huge       (6.55) &  9.5e-12 &    8.77 &  72.00\% &  75.23\% &  76.77\% &  80.00\% \\
              & & Random Forest &    Huge       (3.11) &  8.6e-07 &    8.75 &  84.46\% &  85.84\% &  87.84\% &  88.92\% \\
              & & SVM &   Huge       (11.94) &  3.1e-16 &   23.30 &   0.15\% &   1.23\% &   1.54\% &  12.15\% \\ \cline{2-3}
              & \multirow{2}{*}{Unsupervised heuristic} & RandomWalk@Metric &   Huge       (12.95) &  7.7e-12 &   15.12 &   9.38\% &  20.31\% &  28.12\% &  31.25\% \\
              & & RandomWalk@FI &   Huge       (32.88) &  1.8e-15 &   30.66 &  21.88\% &  21.88\% &  28.12\% &  40.62\% \\ \cline{2-3}
              & \multirow{3}{*}{Ablation Study} & DéjàVu w/o GRU &    Huge       (3.80) &  5.2e-08 &   15.97 &  65.91\% &  70.76\% &  73.18\% &  77.42\% \\
              & & DéjàVu w/o AGG &    Large      (0.95) &  2.4e-02 &    5.77 &  69.70\% &  75.15\% &  78.03\% &  81.06\% \\
              & & DéjàVu w/o BAL &    Small      (0.44) &  1.7e-01 &    5.75 &  62.43\% &  69.39\% &  74.85\% &  78.18\% \\
\cline{1-10}
\multirow{12}{*}{$\mathcal{C}$} & Supervised & DéjàVu &                    - &        - &    1.70 &  61.84\% &  82.63\% &  90.79\% &  96.32\% \\ \cline{2-3}
              & \multirow{2}{*}{Similar failure matching} & JSS'20 &  Huge       (335.31) &  1.5e-24 &   26.34 &  34.21\% &  52.63\% &  60.53\% &  68.42\% \\
              & & iSQUAD &   Huge       (24.53) &  2.5e-14 &    3.50 &  31.58\% &  57.89\% &  63.16\% &  76.32\% \\ \cline{2-3}
              & \multirow{4}{*}{Traditional ML} & Decision Tree &   Huge       (13.20) &  5.4e-17 &   11.21 &  38.95\% &  44.74\% &  45.27\% &  58.42\% \\
              & & Gradient Boosting &   Huge       (24.77) &  7.3e-22 &    3.09 &  44.74\% &  67.37\% &  73.68\% &  77.11\% \\
              & & Random Forest &    Very Large (1.83) &  3.4e-04 &    2.26 &  61.05\% &  75.26\% &  82.11\% &  87.10\% \\
              & & SVM &   Huge       (11.08) &  1.2e-15 &    4.09 &  26.84\% &  46.32\% &  60.26\% &  73.69\% \\ \cline{2-3}
              & \multirow{2}{*}{Unsupervised heuristic} & RandomWalk@Metric &   Huge       (89.17) &  2.3e-19 &    8.25 &  10.53\% &  18.42\% &  34.21\% &  47.37\% \\
              & & RandomWalk@FI &   Huge       (73.25) &  1.3e-18 &    7.08 &  23.68\% &  36.84\% &  42.11\% &  52.63\% \\ \cline{2-3}
              & \multirow{3}{*}{Ablation Study} & DéjàVu w/o GRU &    Large      (0.90) &  3.0e-02 &    2.10 &  63.95\% &  90.00\% &  92.63\% &  96.05\% \\
              & & DéjàVu w/o AGG &    Small      (0.32) &  2.4e-01 &    1.75 &  58.95\% &  88.68\% &  94.48\% &  97.37\% \\
              & & DéjàVu w/o BAL &    Very Large (1.24) &  6.4e-03 &    2.16 &  50.00\% &  74.47\% &  86.84\% &  94.21\% \\
\cline{1-10}
\multirow{12}{*}{$\mathcal{D}$} & Supervised & DéjàVu &                    - &        - &    2.63 &  75.62\% &  84.69\% &  90.42\% &  94.27\% \\ \cline{2-3}
              & \multirow{2}{*}{Similar failure matching} & JSS'20 &  Huge       (529.06) &  4.8e-40 &  303.39 &   8.70\% &   8.70\% &  13.04\% &  18.84\% \\
              & & iSQUAD &  Huge       (451.10) &  4.5e-39 &  259.07 &  21.74\% &  30.43\% &  30.43\% &  31.88\% \\ \cline{2-3}
              & \multirow{4}{*}{Traditional ML} & Decision Tree &   Huge       (24.37) &  5.0e-27 &  128.98 &  64.37\% &  72.19\% &  73.13\% &  73.60\% \\
              & & Gradient Boosting &   Huge       (54.81) &  2.9e-26 &   33.79 &  50.72\% &  59.42\% &  59.42\% &  63.77\% \\
              & & Random Forest &    Huge       (2.65) &  6.4e-07 &   12.35 &  85.66\% &  86.67\% &  89.13\% &  91.88\% \\
              & & SVM &  Huge       (157.82) &  1.1e-32 &   92.35 &   1.45\% &   7.25\% &  10.14\% &  15.94\% \\ \cline{2-3}
              & \multirow{2}{*}{Unsupervised heuristic} & RandomWalk@Metric &  Huge       (179.88) &  1.7e-33 &  104.89 &   2.90\% &   5.80\% &   8.70\% &  13.04\% \\
              & & RandomWalk@FI &  Huge       (776.04) &  2.2e-42 &  443.79 &   0.00\% &   0.00\% &   0.00\% &   1.45\% \\ \cline{2-3}
              & \multirow{3}{*}{Ablation Study} & DéjàVu w/o GRU &    Huge       (2.84) &  2.2e-07 &   13.32 &  68.12\% &  80.94\% &  89.06\% &  91.41\% \\
              & & DéjàVu w/o AGG &    Huge       (2.58) &  9.5e-07 &    3.81 &  83.60\% &  91.40\% &  92.19\% &  93.75\% \\
              & & DéjàVu w/o BAL &    Very Small (0.17) &  3.4e-01 &    2.74 &  72.97\% &  83.13\% &  90.47\% &  94.69\% \\
\bottomrule
\multicolumn{10}{l}{* $\Uparrow$ denotes the improvement rate (\%) of \ours{} over the compared method. Effect sizes (Cohen's d~\cite{cohen1988spa}) and p-values ($t$-test) are calculated based on MAR.}
\end{tabular}

\label{tbl:performance-comparison}    
\end{table*}
\endgroup

In \cref{tbl:performance-comparison}, we present the fault localization results.
We repeat each experiment ten times due to the stochastic training and present the average results.
The MAR of \ours{} achieves 1.66$\sim$5.03 and outperforms the baselines by 11.84\%$\sim$99.41\% on all datasets.
On \ds{B} and \ds{D}, all methods generally perform worse because there are many more candidate failure units (see \cref{tbl:dataset}).
For statistical analysis, we calculate the effect size (Cohen's d \cite{cohen1988spa}) and conduct $t$-test between the MARs of \ours{} and the baselines.
As shown in \cref{tbl:performance-comparison}, the improvement over the baselines is huge and significant ($p {<} 0.05$).
By further calculation, the MAR of \ours{} achieves 2.82 on average and outperforms the baselines by 54.52\%$\sim$97.92\%.
In conclusion, \ours{} achieves good performance and significantly outperforms the baselines.

JSS'20 and iSQUAD perform poorly for two reasons.
First, they do not utilize historical failures until taking the ground truths of the found similar historical failures.
Although carefully designed, their critical intermediate steps (e.g., anomaly detection methods and similarity functions) are completely unsupervised.
Thus, they can be confused by irrelevant abnormal changes in other metrics, which are caused by noises or fluctuation, especially when the number of metrics or failure units is large.
In contrast, \ours{} learns to focus on important metric patterns from historical failures.
Second, JSS'20 and iSQUAD localize faulty failure units by taking the ground truths of similar failures.
However, there may not be a historical failure having the same faulty failure unit as the incoming failure, especially when there are lots of failure units.
In such cases, the faulty failure units' ranks are half of the number of failure units.

The traditional machine learning baselines, on the one hand, perform poorly due to the numerous useless and noisy time-series features.
On the other hand, such traditional machine learning models have poorer capacity than deep learning models, which limits their performance~\cite{goodfellow2016deep}.
As a result, the ensemble models (e.g., Random Forest and Gradient Boosting) with better capacity can perform better than simple models (e.g., Decision Tree).
Though on some datasets (e.g., \ds{B} and \ds{D}), Random Forest achieves the best top-$k$ accuracies, \ours{} still significantly outperforms it with respect to MAR.
It indicates that Random Forest works extremely badly in some cases.
Furthermore, such models could cost too much time in online localization (see \cref{sec:efficiency}).

The two random walk-based methods perform poorly as their underlying intuition (i.e., the correlation of metrics faithfully reflects the probability of failure propagation) could not hold in all situations.
Furthermore, for RandomWalk@Metric, it is hard to construct the causal graph from scratch, especially when the number of metrics is large.
For RandomWalk@FI, there lacks an appropriate method to calculate the correlations of two groups of different metrics.
Finally, as well as JSS'20 and iSQUAD, they are unsupervised and can be affected by noises.

\begin{figure}[htb]
\captionsetup{skip=0pt}
\centering
\includegraphics[width=0.95\columnwidth]{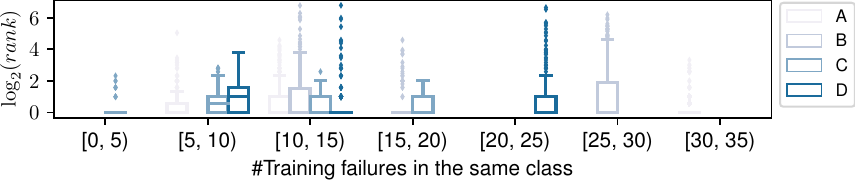}
\caption{Sensitivity to frequencies of training failures}
\Description{The average ranks of failures show no correlation to the number of failures with the same failure class in training data.}
\label{fig:sentivity-to-frequency}
\end{figure}

In \cref{fig:sentivity-to-frequency}, we group testing failures in each dataset by the number of training failures in the same classes (i.e., the failure classes of their faulty failure units) and show the recommended ranks of ground truths of each group.
The results show that \ours{} does not perform worse as the number of training failures in the same class decreases.
Note that the number of outliers differs because the number of failures in each group differs a lot.

In summary, \ours{} is effective in localizing faulty failure units for recurring failures so as to save much effort for engineers and reduce time to mitigate.
Particularly, the performance on real-world failures demonstrates the effectiveness in real-world scenarios.

\subsubsection{Contribution of Main Modules}
\label{sec:ablation-study}
We study the contribution of the main modules used in \ours{} by removing each of them.
In \cref{tbl:performance-comparison}, we compare complete \ours{} with \ours{} without GRU feature extractor (i.e., using 1-D CNN directly, denoted as \ours{} w/o GRU), \ours{} without feature aggregator (i.e., directly feeding the unit-level features into the final classifier, denoted as \ours{} w/o AGG), and \ours{} without class balancing (\ours{} w/o BAL).
In all datasets, the MAR of \ours{} outperforms \ours{} w/o GRU by 19.07\%$\sim$80.25\%, outperforms \ours{} w/o AGG by 3.19\%$\sim$30.92\%, and outperforms \ours{} w/o BAL by 3.85\% $\sim$30.70\%.
By further calculation, on all datasets, the average MAR of \ours{} outperforms \ours{} w/o GRU by 69.03\%, outperforms \ours{} w/o AGG by 20.45\%, and outperforms \ours{} w/o BAL by 15.66\%, and the improvement is significant.
Thus, generally speaking, all three modules contribute significantly to the overall performance.

In particular, there is a small improvement compared with \ours{} \textit{w/o AGG} on \ds{C} because, as introduced in \cref{sec:dataset}, we connect all failure units to a virtual vertex on the FDG of \sys{C}.
The result shows that modeling failure propagation contributes little without meaningful relationships among the failure units.
\ours{} can still achieve high performance in such a case because, in this dataset, the engineers are concerned about the metrics indicating more waiting events and consider little failure propagation.

On \ds{B} and \ds{D}, the improvement over \ours{} w/o BAL is small and insignificant.
It is mainly because in \ds{B} and \ds{D}, the numbers of failures in different classes are close to each other, and there are still many failures in the most minor class.

With respect to top-1 accuracy, the performance of \ours{} keeps similar or slightly poorer than a variant without any one of these modules.
For feature extractor and aggregator, the gradient vanishing problem causes the extracted features to tend to be similar to each other \cite{li2019deepgcns}.
Thus, the features of ground truths get slightly harder to be distinguished in the first place in some cases. 
For class balancing, the modification on training data introduces extra noises and could incur performance degradation.
However, the degradation is slight, and \ours{} always performs well.

In conclusion, the individual modules in \ours{} (GRU feature extractor, feature aggregator, and class balancing) in \ours{} indeed contribute to the overall performance.

\subsection{RQ2: Performance in Various Situations}
\label{sec:configuration-impact}
We investigate how \ours{} performs with varying or incomplete FDGs, insufficient training data, and different model architectures.
Due to the space limit, without loss of much generality, we present results on part of the datasets.  

\subsubsection{Varying and Incomplete FDGs}
\captionsetup{skip=0pt}
\begin{figure}[htb]
\centering
\includegraphics[width=\columnwidth]{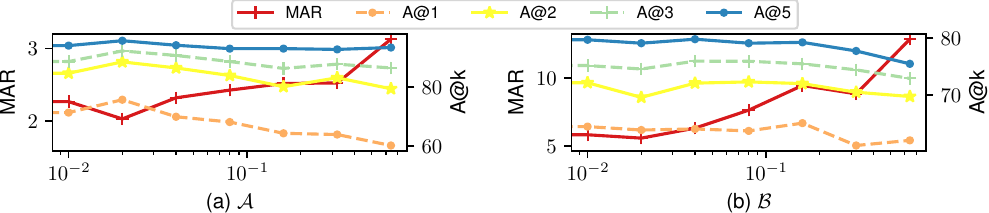}
\caption{Impact of removed FDG edge fractions}    
\label{fig:impact-of-dfe}
\Description{On all datasets, the performance degradation is slight when the fraction of removed FDG edges is not greater than 10\%.}
\end{figure}

The FDGs collected for a system could be incomplete and vary over time due to insufficient knowledge or software/deployment changes.
To evaluate \ours{} under such cases, we randomly remove a fraction of FDG edges for each failure and then train and evaluate \ours{}.
We repeated the experiments ten times.
As shown in \cref{fig:impact-of-dfe}, as the fraction increases, the performance degrades, but the degradation is slight when the fraction is not greater than 10\%.
When too many edges are removed, the performance is worse than \ours{} w/o AGG in \cref{tbl:performance-comparison} because the random FDGs give wrong structural information for feature aggregation.
In conclusion, our approach can handle varying FDGs and achieve relatively good performance with incomplete FDGs.

\subsubsection{Insufficient Training Data}

\begin{figure}[htb]
\centering
\includegraphics[width=\columnwidth]{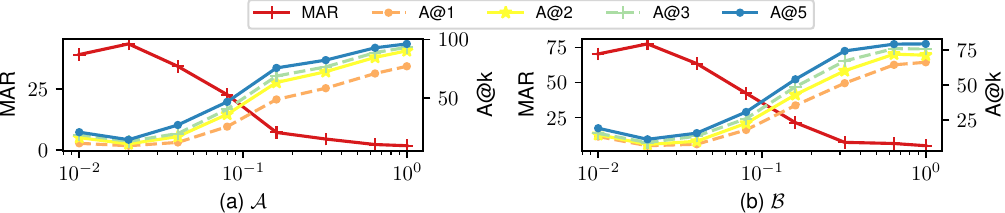}
\caption{Impact of the percent of used training data}
\label{fig:tss-impact}
\Description{As the training set grows, the performance goes better.}   
\end{figure}

In \cref{fig:tss-impact}, we present the performance when only part of the training data is used.
As the training set grows larger, \ours{} performs better in all datasets.
When more than 50\% of historical failures are used, the performance increases slowly.
A possible explanation is that we require only a few training failures for each failure class due to the generalizability of \ours{}.
Thus, it is possible to achieve good enough performance with much fewer training data.

\subsubsection{Different Model Architectures}

\begin{figure}[htb]
\captionsetup{skip=0pt}
\centering
\includegraphics[width=\columnwidth]{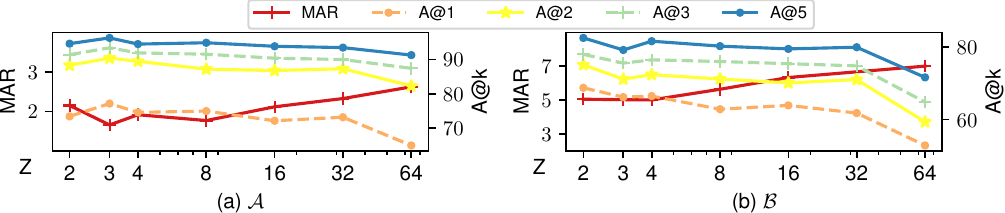}
\caption{Impact of the feature dimension $Z$}
\label{fig:z-dim-impact}
\Description{The performance of \ours{} keeps steady while $Z$ is in a large range.}   
\end{figure}

\begin{figure}[htb]
\captionsetup{skip=0pt}
\includegraphics[width=\columnwidth]{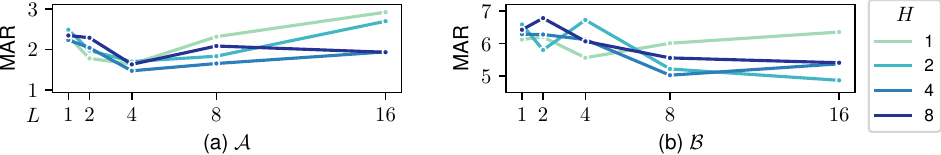}
\caption{Impact of feature aggregator architecture}
\label{fig:aggregator-architecture-impact}
\Description{The impact of feature aggregator architecture}   
\end{figure} 
The model architecture is controlled by $Z$ (the dimension of features), $H$ (\#GAT heads), and $L$ (\#GAT layers).
As shown in \cref{fig:z-dim-impact}, the performance keeps steady as $Z$ changes in a large range.
When $Z$ is small, the feature extractors and aggregators can be considered to use their former layers to extract/aggregate features and use their latter layers to help map the features into suspicious scores.
Therefore, though the feature is of small dimension, it can still keep enough information for the classifier.
In summary, it is acceptable to set $Z$ to our default value (3) when applying \ours{} on other systems.
Moreover, as shown in \cref{fig:aggregator-architecture-impact}, using multiple heads and layers improves the performance compared with vanilla GAT in general.
However, the performance can degrade when H or L is too large.
The best setting of $H$ and $L$ differ in different datasets.
In practice, we set $H$ and $L$ to the default values (4 and 8, respectively), and, when necessary, we can fine-tune them in each online service system for better performance.

In conclusion, \ours{} performs well in various situations.

\subsection{RQ3: Efficiency}
\label{sec:efficiency}
\label{sec:training-time}
\label{sec:localization-speed}

\begin{figure}[htb]
    \captionsetup{skip=0pt}
    \centering
    \includegraphics[width=\columnwidth]{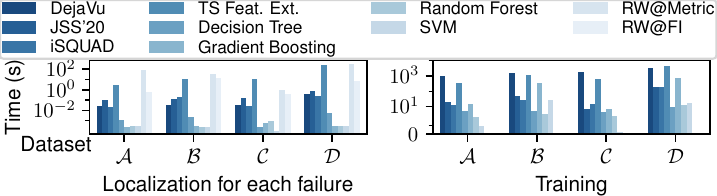}
    \caption{Comparison of time efficiency}
    \label{fig:time-comparison}
    \Description{Time comparison}
\end{figure}

\begin{figure}[htb]
\captionsetup{skip=0pt}
\centering
\includegraphics[width=\columnwidth]{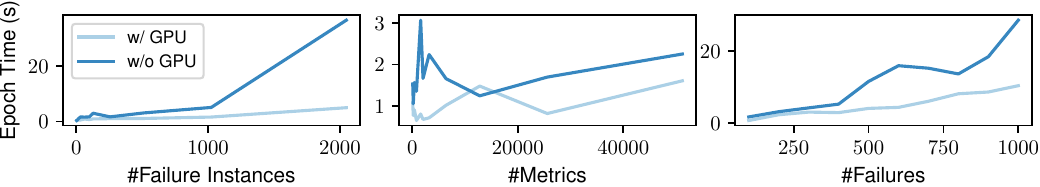}
\caption{Training time for each epoch}
\label{fig:epoch-time-to-factors}
\Description{The training time of \ours{} increases linearly to the number of failure instances, the number of metrics, and the number of failures.}
\end{figure}

We compare the localization and training time with the baselines on servers with Xeon E5-2680 v4, 30G RAM, and GeForce 2080Ti.
As shown in \cref{fig:time-comparison}, \ours{} produces localization results for a failure within one second, which is enough in practice.
Compared with manual localization (several minutes to tens of minutes, see \cref{sec:industrial-localization-practice}), \ours{} can automatically diagnose a failure much more rapidly.
Though the traditional machine learning models are fast, the time-consuming time-series feature extraction (TS Feat. Ext. in \cref{fig:time-comparison}) makes them impractical.
Then, though the training time will not affect the efficiency of online localization, we analyze it to demonstrate the scalability of \ours{}.
As shown in \cref{fig:time-comparison}, though it costs more time to train a \ours{} model than most baselines, the overall time consumption is not large.
Note that the two random-walk methods have no training stage.
In \cref{fig:epoch-time-to-factors}, we further analyze the training time for each epoch with respect to the number of failure instances/metrics/failures with simulated data.
The results show that the training time of \ours{} increases linearly to these factors, and with GPU acceleration, the training time can be significantly reduced.
In conclusion, our approach is efficient and scalable.

\subsection{RQ4: Generalization}
\label{sec:generalization}

\begin{figure}[htb]
\captionsetup{skip=0pt} 
\begin{subfigure}{0.49\columnwidth}
\captionsetup{skip=0pt}                  
\includegraphics[width=\columnwidth]{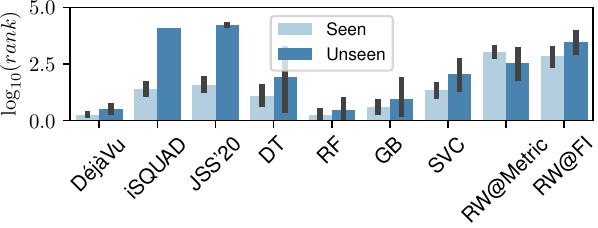}
\caption{\ds{A}}    
\end{subfigure}\hfill
\begin{subfigure}{0.49\columnwidth}
\captionsetup{skip=0pt}
\includegraphics[width=\columnwidth]{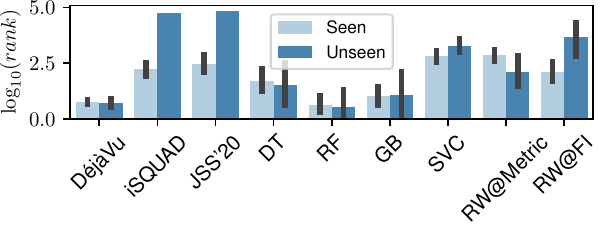}
\caption{\ds{B}}    
\end{subfigure}\hfill
\caption{The recommended ranks of the ground truths of previously seen/unseen failures.}
\label{fig:generalization-compare}
\Description{\ours{} generalizes well in all datasets.}
\end{figure}

We evaluate the generalizability of \ours{} by comparing the performance of previously seen (the ground truths are faulty in some historical failures) and unseen failures.
As shown in \cref{fig:generalization-compare}, for \ours{}, the ranks of ground truths of previously unseen failures are close to those of previously seen failures.
In contrast, iSQUAD and JSS'20, perform significantly worse on previously unseen failures.
It is because they cannot recommend failure units other than those faulty in some historical failures.
A decision tree is shared inside each failure class and thus, can generalize in many cases.
RW@Metric and RW@FI can also generalize since they are heuristic methods and do not rely on historical failures.
However, their overall performance is poor compared with \ours{}.
In conclusion, \ours{} has good generalizability to localize faulty failure units effectively for previously unseen failures.

\subsection{RQ5: Interpretation}
\label{sec:interpretation-evaluation}

\subsubsection{Global Interpretation}
\begin{figure}[htb]
\captionsetup{skip=0pt}
\centering
\includegraphics[width=0.9\columnwidth]{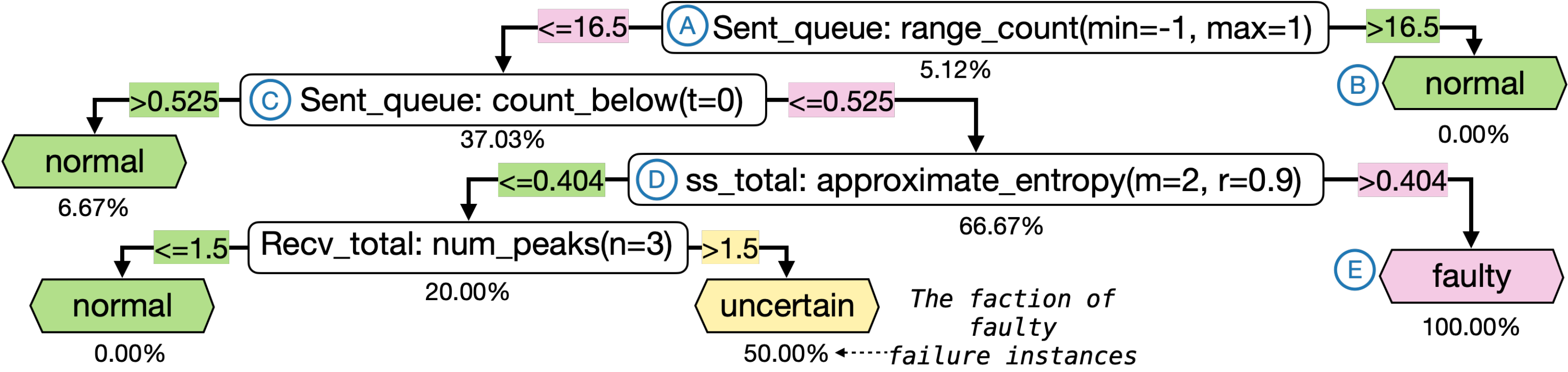}
\caption{A decision tree trained for the failure class \textit{OS Network} of \ds{A}.
Note that the metrics values are normalized.}
\label{fig:global-interpretation-example}
\Description{A example of global interpretation}
\end{figure}

We compare our global interpretation method with two widely-used methods, LR (logistic regression) and LIME~\cite{ribeiro2016why,metzenthin2022lime}.
In \cref{fig:global-interpretation-example}, we present the decision tree given by our interpretation method for \textit{OS Network} failures in \ds{A}.
It shows reasonable rules for discovering network-related issues.
For example, the path \circtext{A}${\to}$\circtext{B} leads to most normal failure units, and its corresponding rule is that range\_count(Sent\_queue, max=1, min=-1)${>}$16.5${\rightarrow}$normal.
When most values of \textit{Sent\_queue} are within the normal range (i.e., [-1, 1]), the send queue is not stuck, and the network works as expected.
The path \circtext{A}${\to}$\circtext{C}${\to}$\circtext{D}${\to}$\circtext{E} leads to most faulty failure units.
The first two conditions (\circtext{A} and \circtext{C}) mean that many values of \textit{Sent queue} are large, indicating the send queue gets stuck.
The third one (\circtext{D}) means that the \textit{ss\_total} is abnormal.
We also confirmed with the engineers that the decision tree is reasonable for this dataset.

\begin{figure}[htb]
\captionsetup{skip=0pt}
\centering
\includegraphics[width=1\columnwidth]{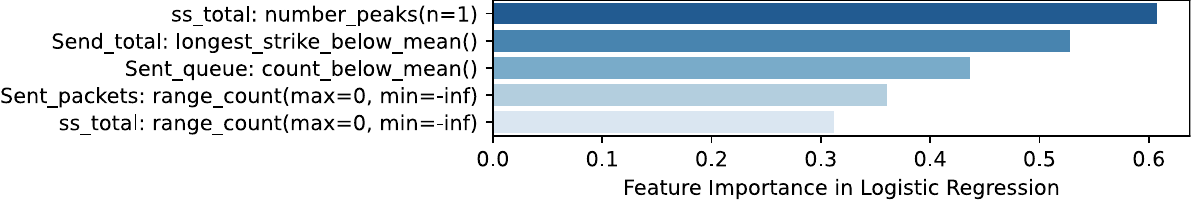}
\caption{The top-5 feature importances given by LR for the failure class \textit{OS Network} of \ds{A}}
\label{fig:lr-global-interpretation-example}
\Description{A example of global interpretation with logistic regression}
\end{figure}

For LR, we train LR models with exactly the same features and targets as the decision trees in our interpretation method and take the coefficients of the features as feature importances.
In \cref{fig:lr-global-interpretation-example}, we present the top-5 feature importances for \textit{OS Network} failures in \ds{A}.
Compared with decision tree, logistic regression cannot give illustrative rules but only importances.

\begin{figure}[htb]
\captionsetup{skip=0pt}
\centering
\includegraphics[width=0.8\columnwidth]{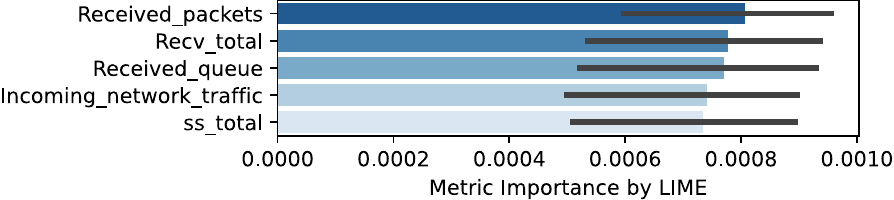}
\caption{The top-5 average metric importances given by LIME on all \textit{OS Network} failures in \ds{A} }
\label{fig:lime-global-interpretation-example}
\Description{A example of global interpretation with LIME}
\end{figure}

LIME interprets individual predictions by learning a local linear approximation of the model to interpret.
In \cref{fig:lime-global-interpretation-example}, we present the average metric importances by LIME on all \textit{OS Network} failures in \ds{A} for global interpretation.
LIME gives only importance scores as well as LR, which are not illustrative.
Furthermore, LIME gives very different metric importances for similar failures, which could be confusing (see \cref{sec:local-interpretation-evaluation}).

\begin{table}
\caption{The decision trees' statistics for global interpretation}
\label{tbl:global-interpretation-statistics}
\footnotesize
\begin{tabular}{cccc}
\toprule
Dataset & Accuracy & \#Nodes & \#Layers \\ 
\midrule
$\mathcal{A}$ & 94.45\% & $5.0\pm 7.1$ & $2.3\pm 2.2$\\
$\mathcal{B}$ & 98.70\% & $49.3\pm 10.0$ & $9.9\pm 11.0$\\
\bottomrule
    
\end{tabular}    
\end{table}
In \cref{tbl:global-interpretation-statistics}, we further present some statistics of the decision trees for global interpretation in dataset \ds{A} and \ds{B}.
The relatively high accuracies show that the decision trees can mimic the deep-learning models well.
The sizes (i.e., the number of nodes and layers) of the decision trees depend on the dataset scale.
The large standard deviations indicate they also vary in different failure classes.

\subsubsection{Local Interpretation}
\label{sec:local-interpretation-evaluation}

\begin{figure}[htb]
\centering
\includegraphics[width=\columnwidth]{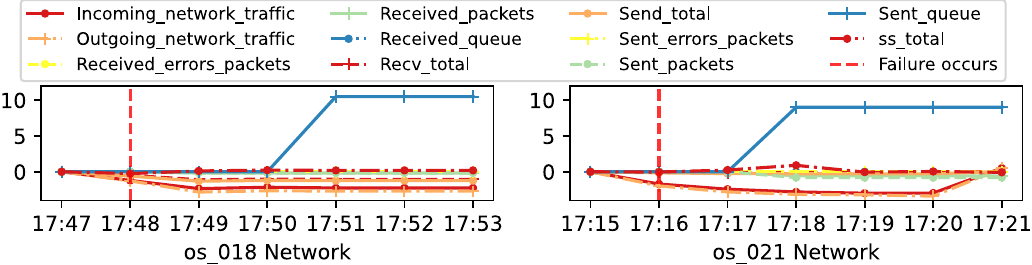}
\caption{
The normalized metrics of the ground truth of a failure (left) and its top-1 similar failure (right).
}
\label{fig:local-interpretation-example}
\Description{A example of local interpretation}
\end{figure}

\begin{figure}[htb]
\captionsetup{skip=0pt}
\centering
\begin{subfigure}{0.4\columnwidth}
\includegraphics[width=\columnwidth]{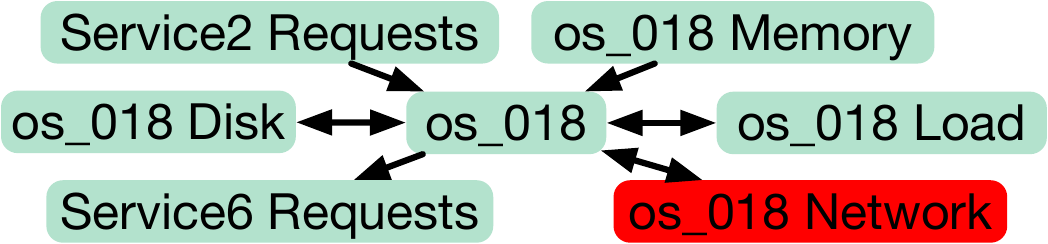}
\end{subfigure}\qquad
\begin{subfigure}{0.4\columnwidth}
\includegraphics[width=\columnwidth]{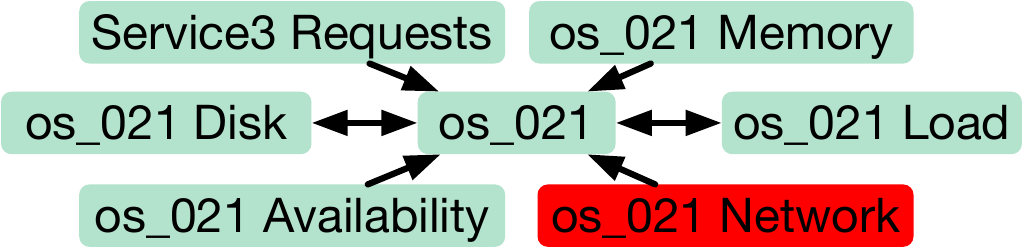}
\end{subfigure}
\caption{The local interpretations given by GNNExplainer}
\label{fig:gnnexplainer-local-interpretation-example}
\Description{A example of global interpretation with GNNExplainer}
\end{figure}

\begin{figure}[htb]
    \captionsetup{skip=0pt}
    \centering
    \includegraphics[width=\columnwidth]{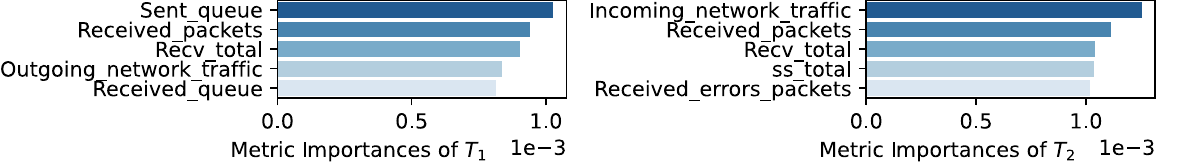}
    \caption{The local interpretations given by LIME}
    \label{fig:lime-local-interpretation-example}
    \Description{The local interpretations given by LIME}
\end{figure}

We compare our local interpretation method with GNNExplainer~\cite{ying2019gnnexplainer}, a state-of-the-art model-agnostic interpretation method for graph neural networks, and LIME.
In \cref{fig:local-interpretation-example}, we present a failure of \textit{OS Network} in \ds{A} ($T_1$) and its top-1 similar failure found by our method ($T_2$).
By presenting $T_2$ to engineers, engineers can understand that we recommend \textit{os\_018 Network} for $T_1$ because its failure scenario is similar to that of $T_2$ in \ours{}'s perspective.
With the help of the failure ticket of $T_2$, engineers could easily understand the underlying failure scenario of $T_1$ and find correct mitigation process.
As a result, the time to diagnose and mitigate can be saved.
For comparison, we present the interpretations of $T_1$ (left) and $T_2$ (right) from GNNExplainer in \cref{fig:gnnexplainer-local-interpretation-example}.
GNNExplainer can only show engineers the important neighbors (green) of the ground truth (red), which requires further efforts to discover why they are important.
In \cref{fig:lime-local-interpretation-example}, we present the top-5 metric importances of $T_1$ and $T_2$ given by LIME.
LIME gives very different metric importances for the two similar failures, making engineers confused.

In summary, \ours{} provides more helpful interpretation.

\section{Discussion}
\label{sec:discussion}

\subsection{Lessons Learned}
\label{sec:lessons}
We summarize some lessons learned from our industrial experience in \ccb{}.
First, the current industrial practice of fault localization relies on the hard-earned intuition from engineers' diagnosing experience, inspired by which we explore learning from historical failures in this paper.
The superiority of \ours{} over the heuristic or unsupervised methods demonstrates that learning from historical failures is effective for fault localization.
Second, periodical re-training is necessary in practice due to the frequent software and deployment change.
In \cref{fig:workflow}, we propose re-training the \ours{} model after mitigating a new failure, which costs only tens of minutes at most each time.
Besides, \ours{} supports using a new FDG, even with new failure units, for each new failure.
Finally, in practice, there are also many non-recurring failures.
Since they have not occurred previously, the suspicious scores of the recommended root causes by \ours{} would be small.
In such cases, we will warn engineers of the potential non-recurring failures.

\subsection{Limitations and Future Works}
\label{sec:limitations}

Not only the faulty metric names but also their patterns are helpful in determining which kind of failure occurs.
Thus, we can integrate metric patterns in the definition of failure units in the future.

Engineers usually expect failure localization solutions compatible with their \textit{existing knowledge}.
Currently, engineers can integrate it into \ours{} by labeling ground truths of historical failures or manually specifying FDGs and failure units.
However, such integration is neither flexible nor accurate enough for all kinds of diagnosis knowledge or rules, which is also a limitation of \ours{}.

In different online service systems, engineers may define completely different failure units, so currently, we cannot transfer a trained model to another system.
However, \textit{transfer learning} in similar systems could be our future work.

The frequent changes in modern online service systems make concept drifts common. In such cases, without enough new failures, we may re-train the DejaVu model by analyzing the metrics' concept drift pattern and adapting the historical metrics accordingly.

\subsection{Threats to Validity}
The \textit{internal} threat to validity mainly lies in our implementation. 
To reduce it, our implementation is based on mature frameworks (\cref{sec:implementation}) and is carefully checked and tested.

The \textit{external} threat to validity mainly lies in the study subjects.
We evaluate \ours{} on four systems, which cannot represent all online service systems.
But we believe our approach is general enough for two reasons.
First, the four systems are from different companies and have completely different architectures.
Particularly, \ours{} is even applicable to traditional software systems (Oracle database for \ds{C}).
Second, our approach is applicable to other online service systems, as the input data of it (e.g., metrics, historical failures, and the relationships between components) are common, and \ours{} is scalable as discussed in \cref{sec:efficiency}.
In the future, we will reduce this threat by evaluating \ours{} on more systems.

The \textit{construct} threat to validity mainly lies in the hyperparameters and evaluation metrics.
We tuned the hyperparameters for \ours{} and baselines by grid-search following existing works \cite{chen2019continuous}.
We also use widely used evaluation metrics (see \cref{sec:evaluation-metrics}).

\section{Related work}
\label{sec:related-work}

\textbf{Fault localization.}
Most existing fault localization methods \cite{wu2021identifying,liu2020unsupervised,cai2021modelcoder,guo2020graphbased,kim2013root} are unsupervised and heuristic.
For example, MonitorRank \cite{kim2013root} takes the historical and current metrics as its input and ranks all root cause candidates with random walk strategy on call graphs.
Prior works also use historical failures to find similar historical failures~\cite{bodik2010fingerprinting,pham2017failure,ma2020diagnosing,brandon2020graphbased} or train supervised localization models~\cite{zhou2019latent,liu2019fluxrank,dogga2022revelio}.
For example, MEPFL \cite{zhou2019latent} treats faulty microservice localization for traces as a multi-class classification problem and directly builds supervised machine learning models (e.g., Random Forest \cite{breiman2001random}) for it.
FluxRank~\cite{liu2019fluxrank} ranks suspicious metric digests in a supervised manner, which are a group of similar metrics and cannot indicate the fault type as failure units do.
Though not directly localizing root causes in a supervised manner, some prior works \cite{ma2020automap} utilize historical failures to tune parameters.
Other than metrics, logs~\cite{nandi2016anomaly,lin2016log,he2018identifying} and traces~\cite{zhou2019latent,li2021practical,yu2021microrank} are extensively studied for fault localization by many prior works.

\textbf{Interpretability.}
Most existing fault localization methods~\cite{zhou2018fault,brandon2020graphbased,bodik2010fingerprinting,pham2017failure,dogga2022revelio,liu2019fluxrank} do not provide interpretation explicitly.
iSQUAD~\cite{ma2020diagnosing}, Fingerprint~\cite{bodik2010fingerprinting}, and JSS'20~\cite{brandon2020graphbased} work by finding similar historical failures.
As discussed in \cref{sec:global-interpretation}, many existing interpretation methods for deep-learning models~\cite{zeiler2014visualizing,bau2017network,dicicco2019interpreting} focus on understanding the inner model mechanism and do not meet our requirements.

\textbf{Deep learning-based program debugging} aims to localize the faulty code elements by deep learning techniques with various features, such as spectrum-based suspiciousness, mutation-based suspiciousness, and complexity-based fault proneness~\cite{li2019deepfl,yang2021survey,lam2017bug,huo2021deep}.
In contrast, the faulty failure units we are concerned about could be either software bugs (e.g., when the components are services), deployment issues, or hardware failures.

\section{Conclusion}
\label{sec:conclusion}
This paper proposes an actionable and interpretable fault localization approach, \ours{}.
The large prevalence of recurring failures motivates us to use supervised models to learn fault localization from the numerous historical failures.
We design a novel deep learning-based model based on graph attention networks, which achieves good performance for this limited scenario (i.e., recurring failures).
To be actionable, our model aims to localize faulty locations and types.
To be interpretable, we propose two interpretation methods.
An extensive study on four systems demonstrates the effectiveness (the average MAR of \num{1.66}$\sim$\num{5.03}) and efficiency (less than one second localization time for one failure) of \ours{}.
The results also confirm the contributions of the main modules in \ours{}.
Particularly, the results on production systems and real-world failures demonstrate our practical performance.

\begin{acks}
This work is supported by the National Key R\&D Program of China 2019YFB1802504, and the State Key Program of National Natural Science of China under Grant 62072264.
\end{acks}

\clearpage{}
\bibliographystyle{ACM-Reference-Format}
\balance
\bibliography{refs.bib}

\end{document}